# Detection of strong light-matter interaction
# in a single nano-cavity with a thermal transducer

Mario Malerba[1,*,#], Simone Sotgiu[2,*], Andrea Schirato[3,4], Leonetta Baldassarre[2], Raymond Gillibert[5], Valeria Giliberti[5], Mathieu Jeannin[1], Jean-Michel Manceau[1], Lianhe Li[6], Alexander Giles Davies[6], Edmund H. Linfield[6], Alessandro Alabastri[7], Michele Ortolani[2,5,&], Raffaele Colombelli[1,5]

[1] Centre de Nanosciences et de Nanotechnologies (C2N), CNRS UMR 9001, Université Paris-Saclay, 10 Boulevard Thomas Gobert, 91120 Palaiseau, France
[2] Department of Physics, Sapienza University of Rome, Piazzale Aldo Moro 5, 00185 Rome, Italy
[3] Dipartimento di Fisica, Politecnico di Milano, Piazza Leonardo da Vinci 32, 20133 Milan, Italy
[4] Istituto Italiano di Tecnologia, via Morego 30, 16163 Genoa, Italy
[5] Center for Life NanoScience, Istituto Italiano di Tecnologia, Viale Regina Elena 291, 00161 Rome, Italy
[6] School of Electronic and Electrical Engineering, University of Leeds, Woodhouse Lane, LS29JT Leeds, UK
[7] Department of Electrical and Computer Engineering, Rice University, 6100 Main Street, TX 77005 Houston, USA

*Equal contribution

ABSTRACT

Recently, the concept of strong light-matter coupling has been demonstrated in semiconductor structures, and it is poised to revolutionize the design and implementation of components, including solid state lasers and detectors. We demonstrate an original nanospectroscopy technique that permits to study the light-matter interaction in single subwavelength-sized nano-cavities where far-field spectroscopy is not possible using conventional techniques. We inserted a thin (~150 nm) polymer layer with negligible absorption in the mid-IR (5μm < λ < 12μm) inside a metal-insulator-metal resonant cavity, where a photonic mode and the intersubband transition of a semiconductor quantum well are strongly coupled. The intersubband transition peaks at λ= 8.3 μm, and the nano-cavity is overall 270 nm thick. Acting as a non-perturbative transducer, the polymer layer introduces only a limited alteration of the optical response while allowing to reveal the optical power absorbed inside the concealed cavity. Spectroscopy of the cavity losses is enabled by the polymer thermal expansion due to heat dissipation in the active part of the cavity, and performed using an atomic force microscope (AFM). This innovative approach allows the typical anticrossing characteristic of the polaritonic dispersion to be identified in the cavity loss spectra at the single nano-resonator level. Results also suggest that near-field coupling of the external drive field to the top metal patch mediated by a metal-coated AFM probe tip is possible, and it enables the near-field mapping of the cavity mode symmetry including in the presence of strong light-matter interaction.

Photonic devices are increasingly developed based on a meta-material approach, where the device functionality stems from the properties of sub-wavelength units/resonators (meta-atoms) that make up the whole optical nanostructured material. The optical response is defined by design obtaining meta-materials [1], meta-surfaces [2] [3], or non-linear surfaces [4] [5] [6] [7]. The unit elements of meta-materials have sub-wavelength dimensions, and they usually operate independently from neighboring units. For this reason, the spectroscopic study of single sub-wavelength units represents an important tool for the development and understanding of such systems: single-object spectroscopy, as opposed to the far-field spectroscopy study of the collective response of large resonator arrays, is enabled by fiber-based or aperture-less scanning near-field optical microscopy (SNOM) [8, 9]. The electromagnetic (EM) near field plays a crucial role in the operation of meta-materials, and SNOM also permits to image near fields.

In the mid-IR and THz spectral ranges, meta-material approaches have recently attracted interest too. On the one hand, the reasonably low ohmic losses at long infrared wavelengths allow the use of metallic meta-structures without introducing excessive damping. This explains why meta-surfaces embedding active materials (detectors [10] [11], lasers/emitters [12], modulators [13]) have recently been employed successfully. On the other hand, the extreme confinement enabled by metal-insulator-metal (MIM) structures permits the study of cavity electrodynamics phenomena, particularly to achieve the regime known as strong light-matter interaction. This is a peculiar condition of interaction between photonic and matter states, of interest not only for fundamental physics [14] [15], but also because of its potential for improving optoelectronic devices in the mid-IR and THz spectral regions [13] [16] [17] [18].

If we consider square patch nano-antennas (a textbook example of an MIM resonant cavity [19]), optical measurements in the far-field require arrays of tens to hundreds of nano-cavities to obtain a good signal-to-noise ratio. There are very few reports of single nano-cavity studies at long infrared wavelengths [20] [21], mostly due to the rather large diffraction limit, and to the inefficiency of mid-IR and especially THz detectors. Furthermore, in patch antennas the core of the architecture – the active region within the optical cavity, where the light-matter interaction takes place – is concealed by the top metal patch and therefore is inaccessible to direct near-field investigation. Optically targeting one single resonator, and simultaneously accessing the concealed active cavity core, thus remains far from straightforward in these conditions.

**RESULTS AND DISCUSSION**

In this work, we present an alternative approach to scattering SNOM [22] [23], based on photo-thermal expansion detection with an atomic force microscopy (AFM) probe [24] [25] [26] [27]. Our approach is capable of detecting the spectral response of a single MIM cavity in the mid-IR range, both in the weak- and strong-coupling regimes. In contrast to SNOM techniques that probe fringing fields leaking outside the cavity and scattered/collected by the probe, our technique is only sensitive to the losses produced by the oscillation of the intracavity field, *via* the generated heating, and can therefore operate even in the ideal case when the electromagnetic field is fully localized inside the cavity, as in MIM resonant cavities. Photothermal simulations reproduce the experimental loss spectra and enable the observed photo-thermal expansion to be linked to the different heat sources in the system. Our technique is also capable of imaging different nanoscale patterns of electric current at the optical frequencies connected to the different resonant modes of the cavities.

The core idea of our approach is summarized in Figure 1: we have fabricated isolated nano-cavities and large arrays of identical nano-cavities on the same chips (Figure 1a) to compare the spectroscopy results of our single nano-cavity technique to those of conventional far-field reflection spectroscopy performed on large arrays. In the inset of Figure 1a a cross-section of the surface layers shows the MIM cavity embedding an active region which, as in traditional studies of strong light-matter coupling regime in the mid-IR, is made of a doped semiconductor quantum well (QW) featuring mid-IR intersubband (ISB) transitions [28]. The ISB transitions provide high oscillator strengths in the mid-IR and THz spectral ranges, and are the backbone principle behind the quantum cascade laser [29] and the quantum well infrared photodetector [30]. ISB transitions are also relevant for studies of fundamental physical phenomena at very long wavelengths. If the light-matter interaction is stronger than the dephasing mechanisms in the system, new quasi-particles – called ISB polaritons – emerge [31] [16]. ISB polaritons have been studied theoretically and experimentally recently, with developments including ISB polariton LEDs [32,33], ultra-fast switching [34], ultra-strongly coupled systems [35,36] and polariton-based detectors [17, 37]. All these results have been obtained with highly confining MIM cavities embedding the active region with an EM overlap approaching unity. The technical innovation introduced here consists in an intra-cavity thin layer of polyethylene (PE) that serves as thermal insulation layer of the active part of the cavity from the heat sink constituted by the substrate, and simultaneously serves as the main thermal expansion layer, thereby amplifying the photoexpansion signal of semiconductor and metal layers above it, which would be too small to be detected by AFM. This large amplification widely compensates the partial loss of EM overlap by a factor of 2 or less between the cavity field and the QW active region. The PE layer acts as a transducer: it receives heat and expands, and such expansion is detected by an AFM. Hereafter, we will refer to our technique as thermal transducer NanoIR (T2NanoIR). A similar technique has been previously employed for much simpler planar circuits in Refs. [8] and [26], but in our work the expansion layer is positioned inside a cavity to detect the heat generated therein. The lateral resolution of heat detection is obviously not limited by diffraction but rather by heat diffusion, which is here carefully engineered to enable spectroscopy of a single nano-cavity (see Supporting Information).

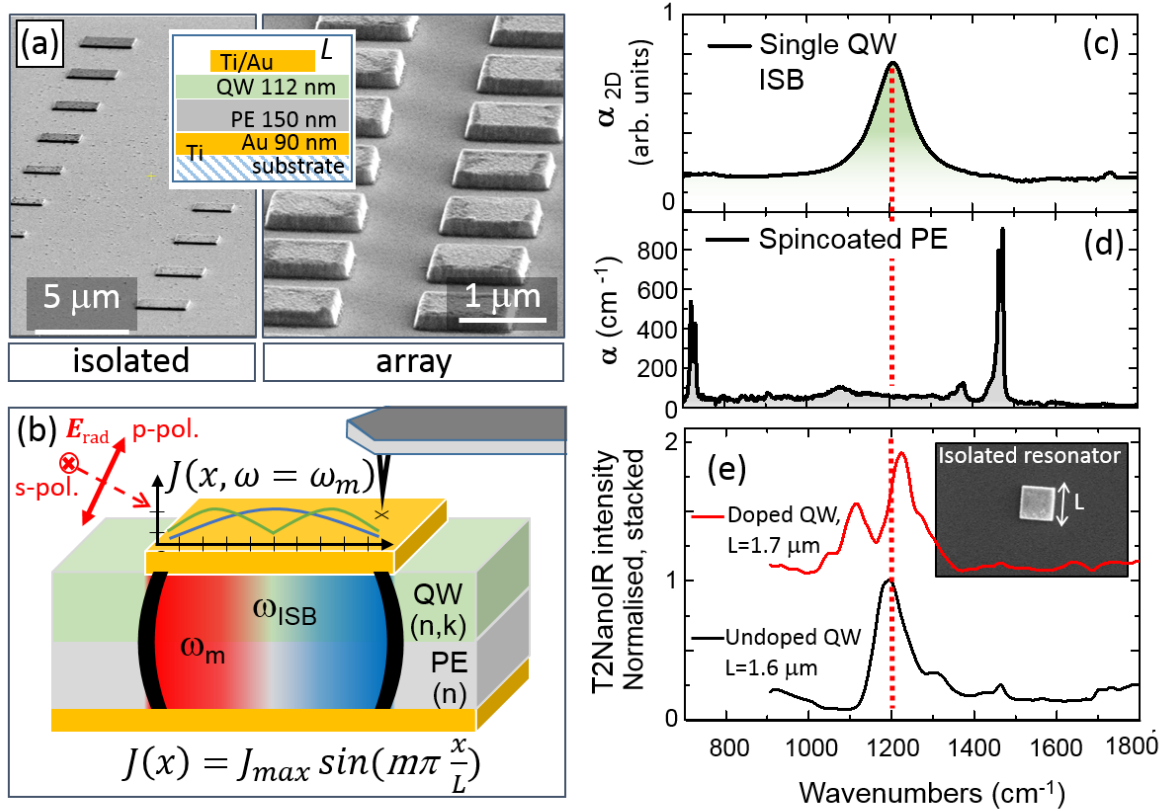

*Figure 1 - **(a)** SEM images taken on different areas of the same chip after device nanofabrication. Left panel: a set of isolated square-patch nano-cavities with different values of lateral side L (the distance between neighboring cavities is > 20L). Right panel: an array of quasi-independent cavities with period P = 2L to be used for far-field spectroscopy. Inset: layer structure of the chip, QW: quantum well; PE: polyethylene; Au: evaporated gold patches and ground plane. The thickness values have been measured from cross-sections obtained by focused ion beam milling. **(b)** Sketch of the T2NanoIR illumination scheme on a single nano-cavity and definition of the main physical quantities involved: $\omega_m$ is the photonic cavity mode resonance frequency; $\omega_{ISB}$ is the intersubband transition frequency in the semiconductor QW; $J_x(\omega_m)$ is the x-component of the electric-current standing-wave pattern on the square patch at the $m^{th}$ mode. **(c)** Absorption spectrum of the doped QW measured by FTIR multipass transmission spectroscopy. The red dotted vertical line highlights the frequency of the ISB transition. **(d)** Absorption coefficient of a PE film spincoated on gold inferred from its FTIR reflectivity. **(e)** T2Nano-IR spectra of single nano-cavities, in the case of an undoped (black curve) and doped (red curve) active region. The former reveals the bare cavity resonance, the latter reveals the two polariton branches.*

In this respect, our technique belongs to the family of techniques that image optical near-fields by detecting non-optical imprints in the material [38, 39, 40]. If inserting a polymer layer in the device structure may appear like a complicated and unnecessary process step, the T2NanoIR technique may be of interest in the context of device physics and technology. In fact, it is not uncommon in the microelectronic industry to plan and implement "dummy wafer" microfabrication runs, in which transistor and diode structures are produced for the sole purpose of studying the thermal dissipation. And besides, the insensitivity of T2NanoIR, if compared to SNOM, to the scattering of the driving laser radiation by nearby photonic integrated circuit elements, including parts and edges of the cavity under study [41], may become a key advantage of T2NanoIR in the field of nanophotonic device characterization (see Supporting Information).

Figure 1b introduces the different quantities involved in the strong light-matter interaction in a single nano-cavity. Metal (Ti/Au) layers are employed for optical confinement. The active region embedded in the MIM resonator is a single InGaAs/AlInAs QW, whose ISB absorption outside the cavity peaks at $\omega_{isb}$~1180 cm$^{-1}$ (Figure 1c). The square patch nano-cavity of side $L$, in the absence of a material with a dipole resonance in its internal volume, resonates at $\omega_m = 4\pi c/nmL,$ where $n$ is the effective refractive index of the dielectric material filling the cavity volume, $c$ is the speed of light and $m$ = 1, 2, ... is an integer mode index. Assuming that the radiation electric field $E_{rad}$ has a nonzero component along $x$, it is possible to infer the electric-current standing-wave pattern that forms on the square patch using the well-known two-curl combination of the third and fourth Maxwell's equations. The electric-current standing-wave patterns $J_x(\omega=\omega_m)$ forming on the resonator for the first two modes m=1 and m=2 are sketched in Figure 1b (for completeness, we mention that 2D modes with simultaneous excitation along $x$ and $y$ are also possible in a two-dimensional square patch cavity, but we do not discuss them in this work). The ~150-nm-thick PE layer is introduced in the cavity using an original nanofabrication technique, which separates the patterning of the active region under study from the realization of the host transducer layer (see Supporting Information). The measured PE absorption coefficient $\alpha_{PE}$ (Figure 1d) does not show any vibrational mode in the frequency range of interest, ideally avoiding the phenomenon of surface-enhanced infrared absorption (SEIRA) that has been exploited e.g. in Ref. [26] for nanoscale imaging. In this sense the polymer acts as a transducer, and not as an active absorber.

When light from a tunable laser source (wavenumber $\omega$) is focused on the sample, it is absorbed at the system resonance and dissipated *via* the different damping channels of the system (ISB absorption and metal ohmic losses). This heats up the material inside the cavity including the PE layer, which features a relatively high linear thermal expansion coefficient and a low thermal conductance that promotes the temperature increase inside the cavity volume $\Delta T$. Scanning $\omega$ while simultaneously measuring the corresponding photothermal expansion with the AFM probe positioned on top of the patch antenna, the cavity loss spectrum can be recorded: it carries information on the single-cavity behavior. For example, in Figure 1e theT2NanoIR spectrum of a cavity loaded with an undoped QW not displaying any ISB transition (black curve) only shows the ohmic losses in the metal patch, which increase when the cavity is at the $m = 1$ resonant mode.

To demonstrate spectroscopy with the T2NanoIR technique on a single nano-cavity, we chose a particularly challenging light-matter interaction condition: the strong coupling regime, where the electromagnetic environment plays a fundamental role in allowing (or preventing) the optical mode to interact strongly with the ISB transition. For the typical polaritonic anti-crossing signature to be detected, the inserted polymer must not induce any important dephasing mechanism or negatively impact the EM overlap in the active region, hampering the coupling strength. Any non-negligible interaction with the polymer would transition the system towards weak coupling, therefore a non-absorbing transducer material must be used. The spectrum in Figure 1e obtained with the T2nanoIR technique on an isolated nano-cavity loaded with an active doped QW (red curve) exemplifies the typical polariton splitting (or anticrossing behavior) observed *in the far-field* in nano-cavity arrays when the cavity resonance matches the absorption resonance of the QW active region.

The signal-to-noise ratio of the T2NanoIR technique can be optimized by employing a gold-coated AFM probe tip in p-polarization at strongly oblique incidence (here 70°, see Figure 1b), a configuration that enhances the radiation field around the tip apex, and by pulsing the laser beam at a repetition rate around $f$ = 150–250 kHz, matching one of the mechanical resonances of the AFM cantilever as discussed in Ref. [25]. Quantum cascade lasers (QCLs) are currently the only mid-IR laser source that can provide single-mode pulsed excitation at any wavenumber *and* at any pulse repetition rate in the range of cantilever mechanical resonances. In this case, the T2NanoIR signal is the spectral component at $f$ of the position-sensitive photodiode signal of the AFM optical lever, detected by lock-in amplification, and it is proportional, also *via* the quality factor of the mechanical resonance (usually around 100), to the actual photoexpansion of the sample under the AFM tip apex $\delta z(\omega)$. In turn, the latter is proportional to the total absorbed radiation power that is the sum of two terms: a direct far-field absorption of $E_{rad}$ by the illuminated patch and a near-field absorption mediated by the enhanced fields around the gold-coated AFM tip. The dependences of $\delta z(\omega)$ on the exact position $(x,y)$ of the tip above the patch, on the laser polarization ($p$ or $s$), and on the AFM tip-coating material (gold or dielectric) is found to be rather weak in the present case, indicating that near-field coupling through the tip is not the dominant energy transfer mechanism from the laser beam to the cavity. Therefore, the T2NanoIR spectra have been acquired by keeping the AFM tip at a fixed position at the center of the square patch, while scanning the wavenumber $\omega$ of the QCL beam. In our experiment, near-field coupling is unessential for single nano-cavity spectroscopy, but nevertheless it is present and we will come back to this point at the end of the paper.

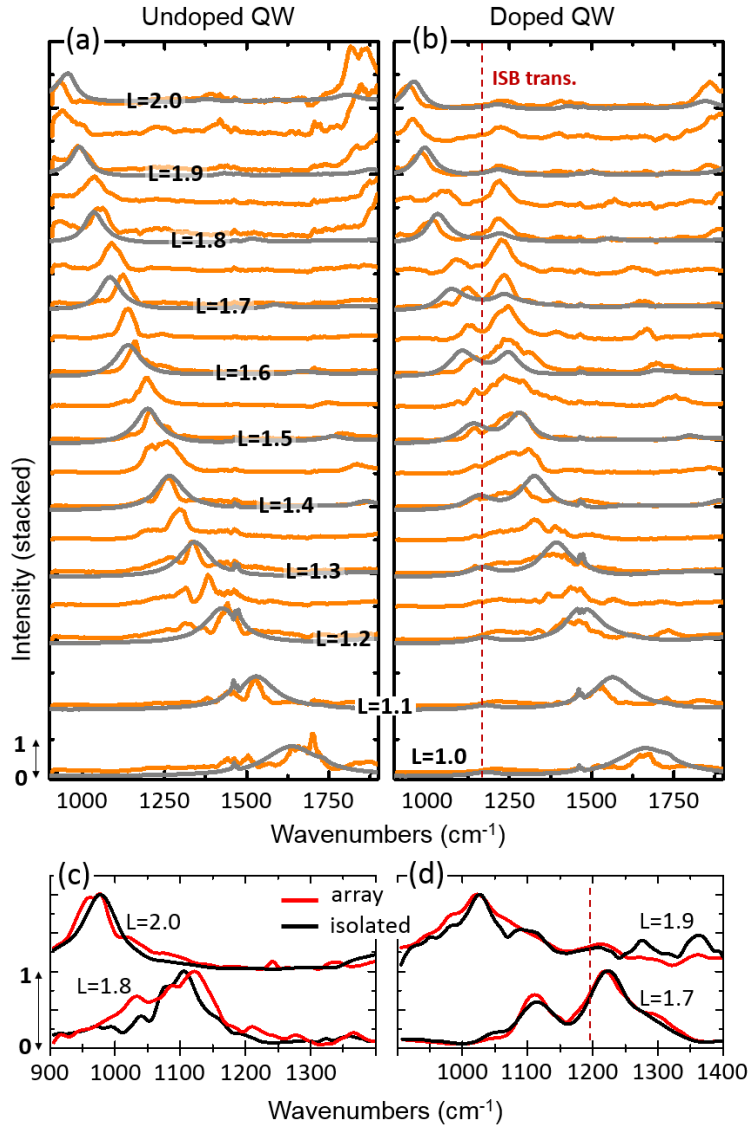

Having demonstrated in Figure 1e that high-quality cavity-loss spectra can be recorded on isolated nano-cavities even in the strong light-matter interaction regime, we now turn to the comparison of the T2NanoIR spectra with more conventional far-field absorption spectra $A(\omega) = 1 - R(\omega)$ where the reflectivity $R(\omega)$ is measured with Fourier Transform IR (FTIR) micro-spectroscopy techniques. Signal-to-noise and diffraction limitations require that arrays of at least a hundred of uncoupled identical nano-cavities be fabricated for FTIR analysis.

Figure 2 reports the spectral comparison between absorption spectra in the far-field on an ensemble of nominally identical patch resonators, and T2Nano-IR spectra on a *single* nano-cavity belonging to the ensemble. The samples comprise arrays of square patch nano-antennas, with side dimensions *L* in the range 1.0 – 2.0 μm, fabricated according to the scheme in Figure 1b. For control, the T2NanoIR spectra have been acquired on several nano-cavities for each array labeled with *L* and the difference in spectra have been found to be not significant (see Supporting Information).

*Figure 2 - T2Nano-IR photoexpansion spectra (orange lines) of individual patch cavities compared to FTIR absorption spectra obtained from reflectivity measurements on the entire arrays (gray lines) for different patch side L. (a) Spectra obtained on cavities loaded with an undoped QW, thus exhibiting purely photonic cavity modes. (b) Spectra measured on cavities loaded with a doped (n= $10^{19}$ cm$^{-3}$) QW, exhibiting a polaritonic behavior and an anticrossing at the ISB transition frequency. (c,d) Comparison of T2Nano-IR aborption spectra taken on isolated patch cavities compared to that taken on cavities belonging to large arrays: (c) undoped and (d) doped QWs. The minor spectral differences than can be found, like the one for the undoped L = 1.8 μm cavities around 1040 cm$^{-1}$, can be ascribed to unwanted SEIRA effect from vibrational absorption by chemical impurities in the PE layer or by nanofabrication process residues.*

Figure 2a shows the measurements on an *undoped* sample (same active region, but no QW doping). The gray lines are the normalized A(ω) measured in far-field. The T2NanoIR cavity loss spectra (orange curves) reasonably overlap with the corresponding A(ω). In panel (a) the fundamental *m* = 1 cavity mode is visible for all values of *L*, and it correctly blue-shifts with decreasing *L* [42]. In the samples with longer *L* the second order one-dimensional mode with *m* = 2 appears at higher wavenumbers, as expected. The minor differences between T2NanoIR cavity loss spectra and the far-field absorption spectra are not systematic and can be attributed to deviation from the perfect square shape of some of the patches and/or unwanted SEIRA effect due to vibrational absorption by chemical impurities in the PE layer or by lithography process residues. As shown in Figure 2c, the thermal transducer enables the same loss spectrum to be captured on a single cavity, and we can confirm that, in the undoped QW samples of Figures 2a and 2c, $\delta z(\omega)$ is proportional to the ohmic metal losses only.

Figures 2b and 2d report the measurements on the *doped QW* sample (the InGaAs layer is doped with Si to $10^{19}$ cm$^{-3}$): the ISB transition, clearly resolved in the two extremal samples L=1.0 μm and L=2.0 μm, interacts with the fundamental cavity mode producing, at all other *L's*, variations of the cavity loss spectrum from the simple sum of non-interacting ohmic and ISB losses. The vertical dotted line at 1180 cm$^{-1}$ corresponds to the position of the ISB absorption peak of Figure 1c. A clear anti-crossing behavior, with a minimum polariton splitting obtained for *L*=1.6 μm, is observed with both techniques and this is clear proof of the strong light-matter interaction regime, even on single isolated nano-cavities (Figure 2d). A summary of the spectral peak positions is reported in Figure 3: it confirms the overall agreement between far-field (FTIR) and T2NanoIR data for both the undoped QW (Figure 3a) and the doped QW operating in the strong light-matter interaction regime (Figure 3b). In the background the coupled-mode theory calculation of the cavity loss spectrum is plotted in the form of underlaying grayscale image. From Figure 3b we can obtain the vacuum Rabi splitting $2\Omega_{Rabi}$, which is defined as twice the Rabi frequency: $2\Omega_{Rabi}$ = 125 cm$^{-1}$ (3.8 THz or 15 meV). It is exactly the same when calculated from T2NanoIR or from FTIR data.

Once we have demonstrated that cavity loss spectra can be equally obtained from far-field reflection spectroscopy in large arrays and from T2NanoIR spectroscopy on isolated (or single) cavities, we can proceed to determine the origin of the temperature increase ΔT that produces the T2NanoIR signal in both the undoped and doped QW samples. Indeed, the quantity $\delta z(\omega)$ peaks approximately at the same ω as the cavity losses (Figure 3) but its spectral intensity will depend on the specific local value of ΔT in the PE layer, which is determined by the thermal transduction process i.e. by the detailed geometry and thermal conductance of each layer and interface. A finite-element method (FEM)-based model has been developed, accounting for the optical and thermal response of the nano-cavity components (from top to bottom: the Au patch, the active doped QW layer, the PE layer and the Au ground plane, lying on a glass substrate in air).

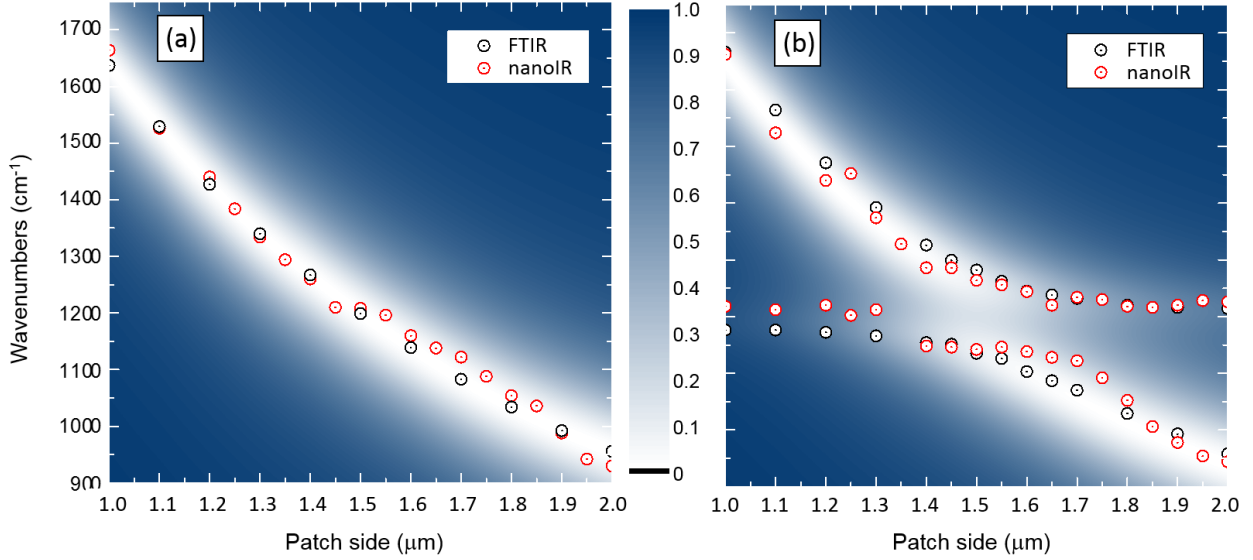

***Figure 3*** *- Cavity resonances peak position as a function of patch side L extracted from the FTIR (black symbols) and T2Nano-IR (red symbols) spectra in fig. 2. Note that FTIR absorption is obtained from an ensemble of identical cavities on 100x100 μm² area while T2Nano-IR photoexpansion peaks are obtained on single cavity resonators. The color map in the background is the expected absorption, calculated analytically by coupled mode theory.* ***(a):*** *undoped QW,* ***(b):*** *doped QW.*

Details of our modelling approach implementation can be found in the Supporting Information. The main results of our numerical simulations are reported in Figure 4 for the two prototypical samples with square patch side $L$ = 2.0 μm (far from the anticrossing point) and $L$ = 1.6 μm (at the anticrossing point). The illumination conditions driving the heat dissipation across the system mimic the T2NanoIR configuration: a $p$-polarised Gaussian beam focus with full width of 25 μm (10 μm) in the x (y) direction and 800 μW of average power impinges on the sample with an angle of incidence of 70°. The optical excitation of the cavity produces a dissipation of the electromagnetic power within the different (lossy) components of the structure $Q = \frac{1}{2}Re\{\boldsymbol{E} \cdot \boldsymbol{J_D^*}\}$, with the local electric field $\boldsymbol{E}$ inducing a local displacement current density $\boldsymbol{J}_D$. The local $Q$ is distinctly defined in each domain of the numerical FEM geometry, and its volume integral determines the amount of power that promotes heating of the sample, however in order to calculate the local ΔT the heat diffusion problem has to be solved. This has been done here for the case of the large nano-cavity array and not for isolated nano-cavities for two reasons: (i) we want to be able to compare the simulations with the broader dataset of Figure 2; (ii) the local ΔT in the central part of the illumination area of a large array can be considered approximately homogeneous in ($x,y$) (see Figure 4a) and this simplifies the interpretation of the thermal simulations in terms of *z*-profiles only. Considering the dimensions of the Gaussian beam, a finite array made of 26x10 patches has been solved for the heat diffusion problem.

Figure 4a shows the calculated spatial distribution of ΔT across the structure for a monochromatic excitation at ω = 970 cm⁻¹; an in-plane (x,y) map is shown along a cross-section of the sample taken at the centre of the QW thickness. Numerical simulations confirm that the (x,y) temperature distribution follows from the laser intensity profile, as shown in the right panel of Figure 4a. Yet, for our investigation, the most

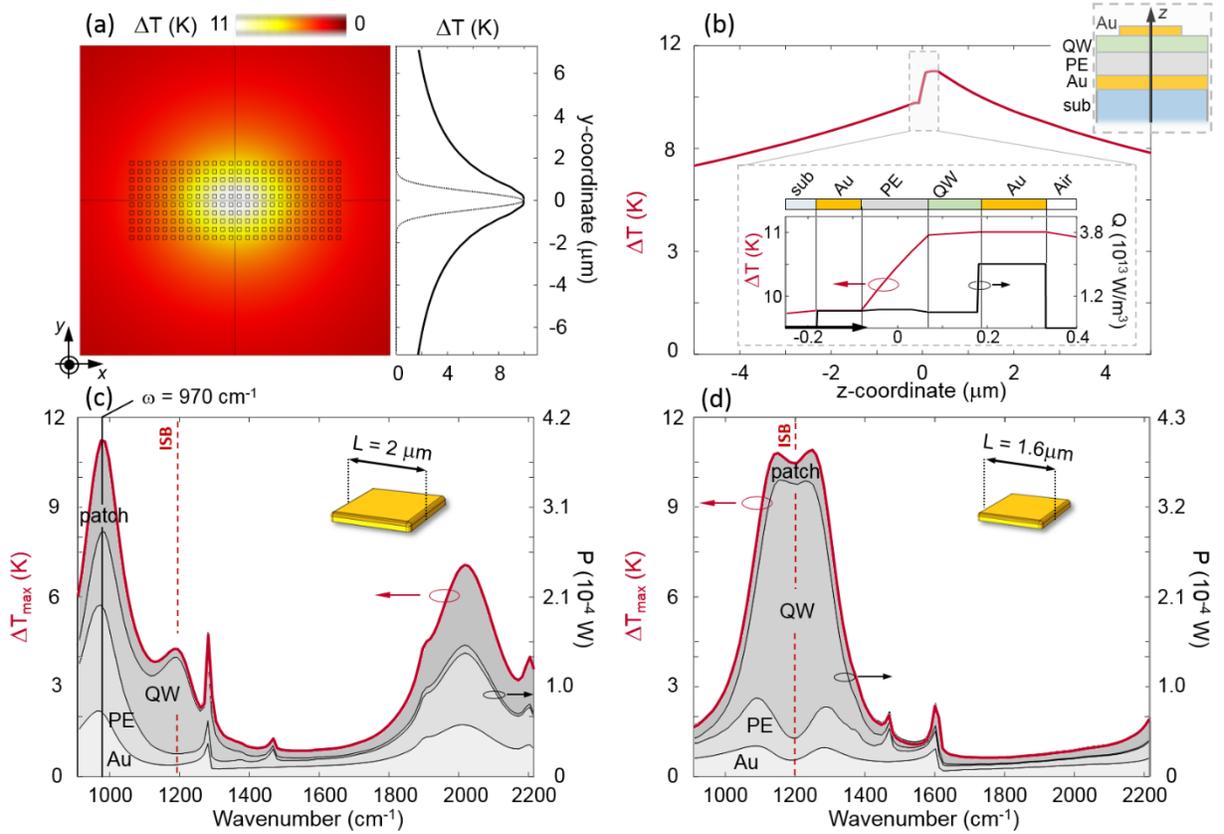

***Figure 4 (a)*** *Spatial distribution of the simulated temperature increase ΔT, shown across a (x,y) cross-section of the sample upon illumination at λ = 10.31 μm (Gaussian beam parameters considered as in the main text, patch side L = 2 μm). In the inset the spatial profile of the temperature increase within the QW layer is evaluated as a vertical cut of the 2D map at x=0, as a function of the y-coordinate (thick solid line), and is compared to the laser intensity profile (thin dotted line).* ***(b)*** *Spatial profile of the simulated temperature increase ΔT evaluated along the vertical line (z) that runs across the device structure on a mm range, i.e. from the silica substrate to the air above of one of the central patches (see top-right sketch). The panel inset shows (red line, left axis) a zoom-in of the temperature increase across the nanocavity structure only. The power density Q (in W/m³) dissipated in each layer is also shown (black curve, right axis) along the same vertical line.* ***(c)*** *Spectrum of the simulated maximum temperature increase (red-line, left axis) reached within the active region for a sample with the patch side L = 2 μm, as a function of the illumination wavenumber. On the right axis, the total heat power is disentangled in the dissipation contributions due to single heat sources (black lines and shaded areas with relative labels).* ***(d)*** *Same as for (c) for a patch side L= 1.6 μm, when heat dissipation is predicted to be dominated by the contribution arising indeed from the QW (note the much larger relative shaded area ascribed to the power dissipated by the QW).*

relevant information is represented by the *z*-profile of the temperature increase *inside* the cavity. To assess its trend, $\Delta T(z)$ was evaluated along a vertical (z) line cutting through all the layers of the central patch. Results in Figure 4b predict that $\Delta T$ (average value ~10.5K) is relatively uniform inside the cavity (inset, red curve), and has a maximum excursion in the PE layer, as expected by design. The inset of Fig. 4b also shows the terms of heat dissipation density Q (assumed to be homogeneous in (x,y)) evaluated along the same vertical line (refer to black arrows in the panel inset) arising from each layer: the highest dissipated power density term is ascribed to the Au square patch.

Figure 4b also shows the terms of heat dissipation density (assumed to be homogeneous in (x,y)) evaluated along the same vertical line (refer to black arrows in the panel inset) arising from each layer: the highest dissipated power density term is ascribed to the Au square patch.

Heat dissipation arises from the interaction with the electromagnetic field, hence the thermal response of the structure varies with the considered excitation wavenumber and is demonstrated to be sensitive to the character of the light-matter interaction regime, either far from, or at, the anticrossing point. To illustrate this effect, Figures 4c and 4d (red curves, left axis) report the maximum temperature increase evaluated within the active region as a function of the impinging laser wavelength for the two values L=2.0 μm and 1.6 μm. Simulations can be compared to the corresponding T2NanoIR spectra in Figure 2. Figure 4c corresponds to a large detuning between cavity and ISB transition, while in Figure 4d the perfect anticrossing point is reached. In both cases, the simulations well reproduce the experimental spectral positions of the peaks, and their relative intensity (dotted lines in panels c and d). Even the SEIRA-like enhancement of the PE vibrational lines at 1370 and 1460 cm$^{-1}$ is reproduced. On the other hand, simulations slightly overestimate the linewidths, since they were performed without any fit parameters and considered a standard 10% FWHM for the ISB transition.

The photothermal simulations provide a deeper insight into the origin of the dissipation in the system and highlight the hybrid character of the polaritonic modes. In Figure 4c, the black curves (right axis) and relative shaded areas represent the cumulative dissipated power *P* (the power density *Q* integrated over the corresponding component volume), disentangled in terms of contributions arising from the individual heat sources within the different cavity components. Far from the anti-crossing, when *L* = 2.0 μm and the fundamental cavity mode is far from the ISB transition, the interaction is non-resonant and it does not enhance heat dissipation in the QW. On the other hand, at perfect anti-crossing (*L* = 1.6 μm, Figure 4d) the fundamental cavity mode with *m* = 1 resonates with the ISB transition, enabling the anticrossing behavior and amplifying the heat dissipation in the QW, which interestingly dominates over ohmic dissipation in the metal layers (compare shaded areas in Figure 4d). We also notice that dissipation in the PE layer cannot be considered as totally negligible, due to residual vibrational absorption strength in the polymer and high field enhancement in the cavity. Obviously, using a polymer with vibrational resonances in this wavelength range would lead to a very large SEIRA effect that would obscure any other thermal behavior in the cavity.

In the final part of the work we investigate if we can resolve different nanoscale fields and/or current patterns connected to the different cavity modes using the T2NanoIR in imaging mode, i.e. acquiring the T2NanoIR signal with the QCL wavelength fixed at the resonant frequency $\omega_m$, while scanning the AFM tip position (x,y) on the surface of the top metal patch. Given the homogeneity of $\Delta T(x,y)$ for each excitation condition discussed in Figure 4, T2NanoIR in imaging mode provides a measurement of the near-field-coupled energy as a function of the gold-coated AFM tip position (x,y). We start from a specific undoped individual resonator with patch side $L$=2.0 μm whose T2NanoIR spectrum in Figure 2(a) shows the m = 1 mode at 970 cm$^{-1}$ and the m = 2 mode at 1850 cm$^{-1}$. We thus perform T2NanoIR imaging at $\omega_1$ and $\omega_2$. A subset of the pixels inside the patch perimeter has been selected for imaging, corresponding to 90% of the patch area, to avoid measurement artefacts stemming from ascending/descending topography. We obtain the maps in Figs. 5a and 5b for $\omega_1$ and $\omega_2$, respectively. Red pixels correspond to high signal. The S/N ratio is low due to the strong (x,y)-independent far-field photoexpansion signal discussed in the paper.

Nevertheless, we clearly observe a single lobe at $\omega_1$, and two lobes with a nodal line in the center at $\omega_2$. The one-dimensional (1-D) symmetry of the lobes allows one to sum, over the orthogonal direction $y$, the T2NanoIR image profiles along $x$ hence obtaining the 1-D intensity profiles shown as plots above the maps. These profiles are reminiscent of a 1-D standing wave pattern corresponding to the electric-current standing waves $J(x)$ in the top metal patch sketched in Figure 1b. Interestingly, in a one-dimensional cavity approximation, the application of the fourth Maxwell's equation $\nabla \times \boldsymbol{H} = \boldsymbol{J} + \varepsilon_0 \partial \boldsymbol{E} / \partial t$ shows that the electric-current patterns have the same symmetry as the pattern of the in-plane magnetic field components $H_y$. These can be easily extracted from EM simulations and are reported in Figures 5(a) and 5(b). On the other hand, the vertical electric field $E_z$, which is usually taken as the dominant near-field interaction quantity e.g. in SNOM experiments [39], would have $m$+1 lobes instead of $m$ lobes as it can be inferred from the third Maxwell's equation $\nabla \times \boldsymbol{E} = -\mu_0 \partial \boldsymbol{H} / \partial t$ (see Supporting Information for all the simulated field intensity patterns).

From this observation we suggest that the metal-coated AFM-tip excites the electric current $J(x)$ on the patch surface through the near-field enhanced component $E_x$, instead. The excitation of an electric current along x is possible in $p$-polarization because, at 70° incidence, there is still a minor component of the radiation electric field directed along x, and the electric field enhancement is certainly asymmetric along x [43], leading to a net charge displacement in one direction as a function of time (see sketch in Figure 5(e)). The near-field enhancement of $E_x$ can be taken as position-independent within the square patch area far from the edges, but the available effective conductivity of the patch depends on x through the standing wave relation $J(x)=J_{max}\ sin(m\pi x/L)$. In this respect, the electric current standing wave pattern behaves like a local density of optical states (LDOS) usually defined in scattering-SNOM problems [44], although this is only a similarity and the two concepts are well separated.

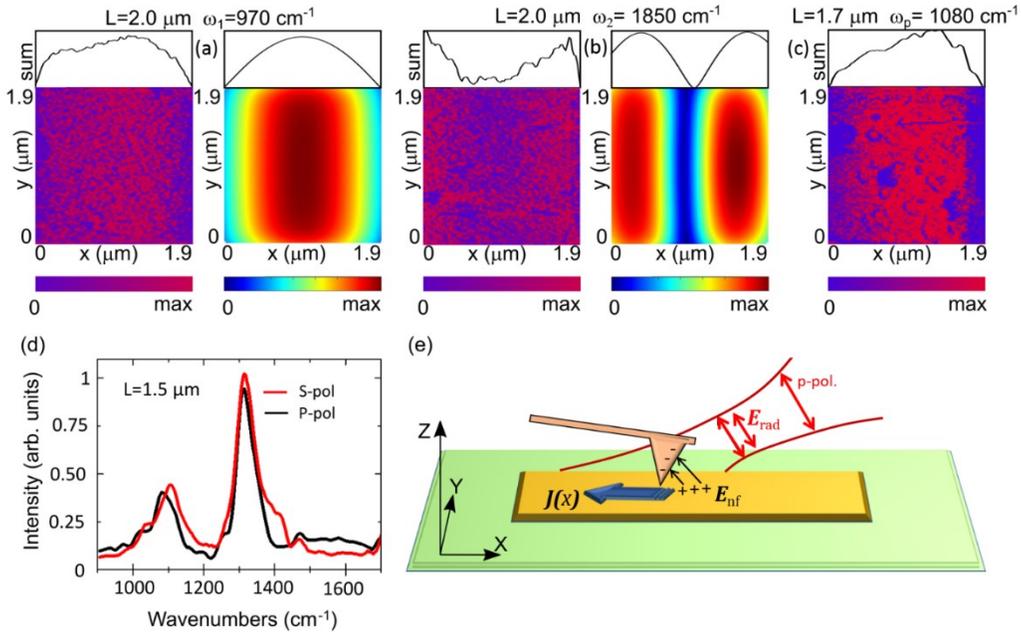

**Figure 5** – Near-field imaging of nanoscale patterns on the patch resonator with side L = 1.9 $\mu$m top surface for different cavity modes. **(a)-(b)** T2NanoIR signal maps (left) and calculated in-plane magnetic field from FDTD simulations (right) at frequencies matching the m=1 **(a)** and m=2 mode **(b)** respectively. The top insets are summed line profiles calculated from the corresponding maps. Note that in 1D approximation the in-plane magnetic field $H_y$ patterns correspond to the electrical current patterns, therefore the mode symmetry is the same for $H_y$ and $J(x)$. Also note that a constant background has been subtracted from the T2NanoIR maps. **(c)** T2NanoIR signal map taken for a patch resonator with side L= 1.7 $\mu$m (thus fully included in the map) for the lower polariton peak frequency $\omega$ = 1080 $cm^{-1}$: a single lobe pattern is observed as for the unperturbed m = 1 cavity mode. **(d)** Comparison of T2Nano-IR spectra acquired on the same patch without moving the AFM tip but rotating the polarization from p to s thereby eliminating any near-field enhancement by the tip. Due to the square shape of the patch, there is a 90 degrees rotation invariance. The antenna spectrum therefore does not depend on the linear polarization direction: only the action of the tip could have made a difference, but this was not the case. **(e)** Sketch of the near-field coupling mechanism generating the image contrast in p-polarization. According to this view one can picture the experimental maps in **(a)**, **(b)** and **(c)** as maps of the near-field coupling strength.

In Figure 5(c) we perform a similar mapping analysis of a doped-QW patch close to the anti-crossing point with L=1.7 $\mu$m. We use a laser wavenumber $\omega_{pol}$ = 1080 $cm^{-1}$ corresponding to the lower polariton peak frequency, which, because of the anti-crossing behavior, is frequency-shifted with respect to the *m*

= 1 mode of the corresponding undoped QW around 1180 cm$^{-1}$. Nevertheless, the electric-current standing-wave symmetry of the lower polariton resonance mirrors the one-lobe symmetry of the m = 1 fundamental mode of the undoped QW patch of Figure 5(a). This behavior can be easily understood even in a semiclassical framework for polariton splitting theory, where the ISB transition is modeled as a Lorentzian resonance that leads to a strong modification of the value of the effective refractive index from the quasi ω-independent value of the undoped QW [42]. In other words, the one-dimensional mode equation $\omega_m = 4\pi c/nmL$ also holds for polariton modes if a suitable effective index is used in substitution of the refractive index of the dielectric-loaded cavity.

As anticipated, in the present case of strong light-matter interaction in square patches, the magnitude of the near-field tip-coupled energy is much smaller than the energy directly coupled by the free-space laser radiation into the patch. The spectral efficiency of the tip-coupling mechanism is almost flat, as further demonstrated in Figure 5(d) by the two T2NanoIR spectra taken on the same patch (L = 1.5 μm shown as an example) and without moving the gold-coated AFM tip by simply rotating the polarization from *p* to *s* through a periscope. Changes seen in Figure 5(d) are interpreted as minor modifications due to slightly different in-coupling geometry. While a full study of the near-field coupling mechanism is beyond the scope of this paper, our suggested interpretation of the imaging contrast mechanism in terms of excitation of the local electric current standing wave is supported by two initial tests: (i) cavity mode imaging by using a dielectric tip, and (ii) cavity mode *imaging* with s-polarization while still using a metal-coated tip. In both cases, no imaging contrast was observed: indeed, the dielectric tip does not couple to the electric current, and an electric field in s-polarised radiation cannot produce an asymmetric field enhancement along *y* (if any) because it is orthogonal to both the tip axis (z) and the incidence plane (xz). This finding confirms the added near-field mapping capability of T2NanoIR in the case of gold-coated tips and p-polarized laser illumination, which is also the most common configuration for nano-IR spectroscopy instruments. Importantly, it reveals that the T2NanoIR maps do not represent the local heating in each xy position inside the cavity, as in this case we would have observed patterns corresponding to the local $E_z$ field distribution. Instead, they represent a map of the near-field coupling strength between the patch and the tip apex, which is found to be higher where the EM field component $H_y$ (hence the electric current J(x)) is more intense.

## CONCLUSIONS

In summary, we have demonstrated an original technique, the thermal-transducer-NanoIR (T2NanoIR), that permits the study of the light-matter interaction inside single isolated nano-cavities. It relies on the insertion of a non-perturbative polymer transducer in the cavity that permits the detection of the absorbed optical power in the form of thermal expansion driven by optical energy dissipation into heat. The ability of T2Nano-IR to measure optical energy dissipation at the nano-scale, coupled to the predictive character of the thermal simulations, could enable in the future the individual measurement of the contributions of the different dissipation channels, even in complex sample geometries. This could help design optimization: for instance, in detectors solely the losses in the active region yield a signal, while all the other dissipation channels are of course unwanted.

## EXPERIMENTAL METHODS

*Materials.* The samples are doped InGaAs/AlInAs semiconductor QWs epitaxially grown by MBE on low-doped InP substrates. The undoped sample contains three 10.5-nm-thick QWs separated by 15-nm-thick barriers, for a total epitaxial thickness of 151.5 nm, while the doped sample is a single multisubband plasmon quantum well, made of a 18.5 nm thick well, Si-doped to $10^{18}$ cm$^{-3}$ and surrounded by InGaAs barriers for a total thickness of 112 nm. Products for the deposition of polyethylene (HDPE powder and decahydronaphthalene solvent) were purchased by Sigma Aldrich. The full fabrication sequence and methods are described in the first supplementary information section.

*Characterization.* <u>Far-field characterization</u> was carried out using a Thermo Nicolet Nexus 870 FTIR connected to a microscope unit (Nicolet Continuum) equipped with a 32x Cassegrain objective. Illumination was provided with a globar source and measurements performed in reflectivity configuration, detecting the optical response with a cooled MCT detector. <u>AFM-IR measurements</u> were performed using a NanoIR2 (Bruker - Anasys Instruments), purging with dry air both the optics and the sample compartment for many hours. The beam from a broadly tunable mid-IR QCL (MIRCATxB, Daylight Solutions, with a spectral range 900–1800 cm−1) was tightly focused onto the gold-coated tip of an AFM probe. In p-polarization, the IR beam impinges from the side at an angle of 70° to the surface normal, that is, with the electric field oriented at 20° with surface normal. In s-polarization, obtained by inserting a custom reflective periscope between the laser and the NanoIR2 and then using the same optical path, the electric field is parallel to the surface. The laser provided 260 ns long light pulses at a repetition frequency in resonance with the second mechanical bending mode of the cantilever at ~200 kHz (resonantly enhanced infrared nanospectroscopy, REINS). The laser power was adjusted using transmission metal-mesh filters in front of the QCL output and it was in the range 1–50 mW depending on frequency range and sample absorption. The laser spectrum is measured with a pyroelectric detector and used as reference for the incoming laser power.

## SUPPORTING INFORMATION

Supporting Information is available free of charge in a separate document online. The covered contents are: (i) detailed device fabrication steps; (ii) details of the employed simulation model (COMSOL) and extension of calculated results for a highly thermally conductive substrate; (iii) simulated electromagnetic patterns of the photonic cavity modes; (iv) a comparison between SNOM and T2NanoIR measurements, showing the unambiguity of results from the latter; (v) considerations on reproducibility (different devices *and* different active regions).

## ACKNOWLEDGEMENTS

We acknowledge financial support from the European Union FET-Open Grant MIRBOSE (737017). This work was partly supported by the French RENATECH network. We acknowledge financial support from the French National Research Agency, projects IRENA (ANR-17-CE24-00016) and SOLID (ANR-19-CE24-0003); from the Italian Ministry of Research (MIUR project PRIN 2017Z8TS5B); and from the EPSRC programme grant "HyperTerahertz" (EP/P021859/1). M.M. acknowledges support from the Marie Skłodowska Curie Action, Grant Agreement No. 748071. We also wish to thank Dr. Andrea Cattoni for the useful discussions on polyethylene spincoating, Adel Bousseksou for the everyday scientific exchange and the cleanroom staff (Jean-René Coudevylle, among others) for the valuable technical support.

# Detection of strong light-matter interaction in a single nano-cavity with a thermal transducer

Mario Malerba[1,*,#], Simone Sotgiu[2,*], Andrea Schirato[3,4], Leonetta Baldassarre[2], Raymond Gillibert[5], Valeria Giliberti[5], Mathieu Jeannin[1], Jean-Michel Manceau[1], Lianhe Li[6], Alexander Giles Davies[6], Edmund H. Linfield[6], Alessandro Alabastri[7], Michele Ortolani[2,5,&], Raffaele Colombelli[1,5]

[1] Centre de Nanosciences et de Nanotechnologies (C2N), CNRS UMR 9001, Université Paris-Saclay, 10 Boulevard Thomas Gobert, 91120 Palaiseau, France
[2] Department of Physics, Sapienza University of Rome, Piazzale Aldo Moro 5, 00185 Rome, Italy
[3] Dipartimento di Fisica, Politecnico di Milano, Piazza Leonardo da Vinci 32, 20133 Milan, Italy
[4] Istituto Italiano di Tecnologia, via Morego 30, 16163 Genoa, Italy
[5] Center for Life NanoScience, Istituto Italiano di Tecnologia, Viale Regina Elena 291, 00161 Rome, Italy
[6] School of Electronic and Electrical Engineering, University of Leeds, Woodhouse Lane, LS29JT Leeds, UK
[7] Department of Electrical and Computer Engineering, Rice University, 6100 Main Street, TX 77005 Houston, USA

*Equal contribution

**Table of contents**

## I. Device Fabrication

The approach aimed to develop a technique that can be applied to any kind of active region and polymeric transducer material, and that reduces possible contaminations to the polymer layer. Since the optical cavity is a hotspot where the electric field is enhanced, any unwanted processing chemical contamination would lead to spurious absorption peaks, which would yield noisy or unreliable spectral results. To comply with such requirements, we separated the fabrication of the nanoantenna patch antennas on the active region from the realization of the thin expanding polymer layer, reducing the whole process to a sequence of conventional steps.

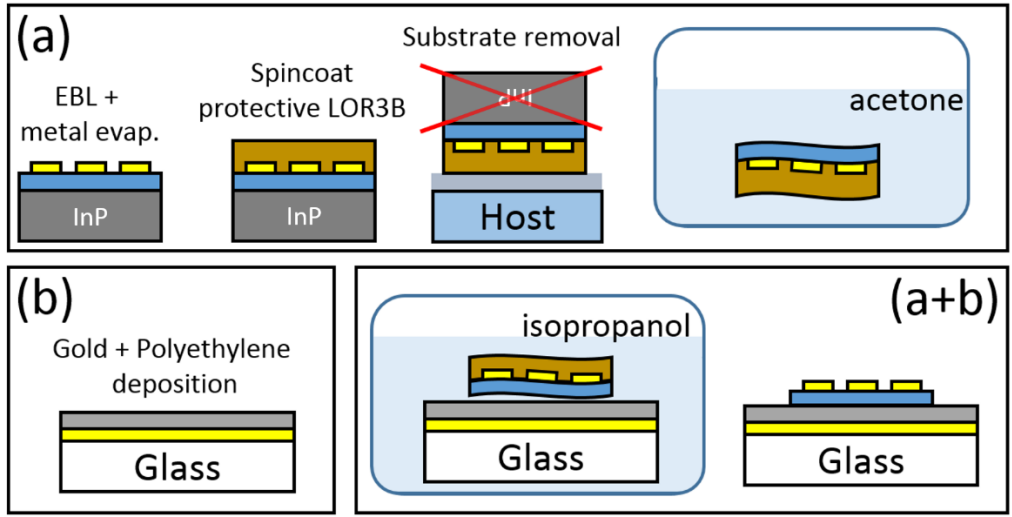

*Figure S 1 Fabrication steps followed to insert a thin layer of pure high-density polyethylene inside a metal-semiconductor-metal optical cavity.*

We first perform electron beam lithography on the active region, grown on its native InP substrate. A layer of metal (10 nm Ti + 90 nm Au) is deposited by electron beam evaporation and lifted off; the sample is then protected with a ~1-μm-thick layer of LOR3B (Microchem) resist, then flipped and temporarily glued face down to a glass slide with a thermolabile, acetone-soluble wax (crystalbond 509). The InP growth substrate is removed in an appropriate etching solution (37% HCl acid), until the etch-stop layer (40 nm of InGaAs) is reached.

In parallel, a 100-nm-thick layer of metal (10nm Ti, 90nm Au) is evaporated on a glass host substrate, and a thin layer of polyethylene is deposited. Polymer deposition is carried out in the following way: HDPE powder (high density polyethylene, purchased by Sigma Aldrich, melting point 140 °C) is dissolved in a solvent with a sufficiently high boiling temperature (decahydronaphthalene, boiling point at 190 °C, pure, also purchased by Sigma-Aldrich). The layer is thus deposited by pre-heating the gold-coated glass support at a temperature of 170 °C, by rapidly transferring it on the spincoater support and by dropcasting and spincoating the support with the hot solution. For a concentration of (100 mg HDPE in 2.5 g decaline), a thickness of 150 nm is obtained with a rotation speed of 2000 rpm. To obtain a uniform film with a deviation of ± 20 nm from average thickness, and to avoid porosity and solvent leftovers, a final annealing of the polymer is performed at its melting temperature for one minute, then rapidly cooling it down at room temperature by dipping it in cold water. Homogeneity is evaluated by observing a lack of interference patterns in the thin film, while the final thickness is evaluated by profilometry.

The patterned active region is transferred onto the host substrate by dissolving the wax in acetone. The heterostructure, which measures approximately 1x1 mm$^2$, protected by LOR3B (not soluble in acetone), is mechanically resistant yet sufficiently flexible to be easily transferred into several clean rinsing baths (acetone and isopropanol), to remove any traces of wax. The final step consists in placing it on the host HDPE/gold substrate, allowing it to dry completely, and removing the protective LOR3B in its developer MF319 (Microchem).

Such an approach guarantees that all fabrication steps can be followed and verified in a repeatable manner, and minimizes contamination or degradation of the thin polymeric layer during the lithographic steps.

## II.    Numerical simulations

To predict the photothermal behavior of the system, a numerical model has been developed using a Finite Element Method (FEM)-based software (COMSOL Multiphysics 5.6), following a segregated approach. First, the model solves for the electromagnetic problem across the sample to assess the dissipated power density ($Q$, in the main text) within each of the domains of the numerical geometry. In light of the periodicity of the patches ($2L$), ensuring a negligible interaction between nearest neighbors, simulations were performed assuming an infinite array of patches. Port formalism was employed and periodic boundary conditions were applied in the $x$ and $y$ directions. Once the electromagnetic problem has been solved, the dissipation power density terms are extracted and averaged over the corresponding domains (a licit assumption, given their substantial locally uniform profiles within the modelled single unit cell of the array). This provides spectra of dissipation power density for the lossy layers of the structure, subsequently used as input data for the thermal simulations. In our case, the Gaussian beam is illuminating a finite number of nanoantennas. Therefore, the thermal simulations are performed over a finite array of patches, with a size estimated from the spot size of the laser beam (exploiting the 4-fold symmetry of the system). Domain heat sources are defined according to the solution of the electromagnetic problem, and infinite element domains are defined at the extreme boundaries of the simulation domain.

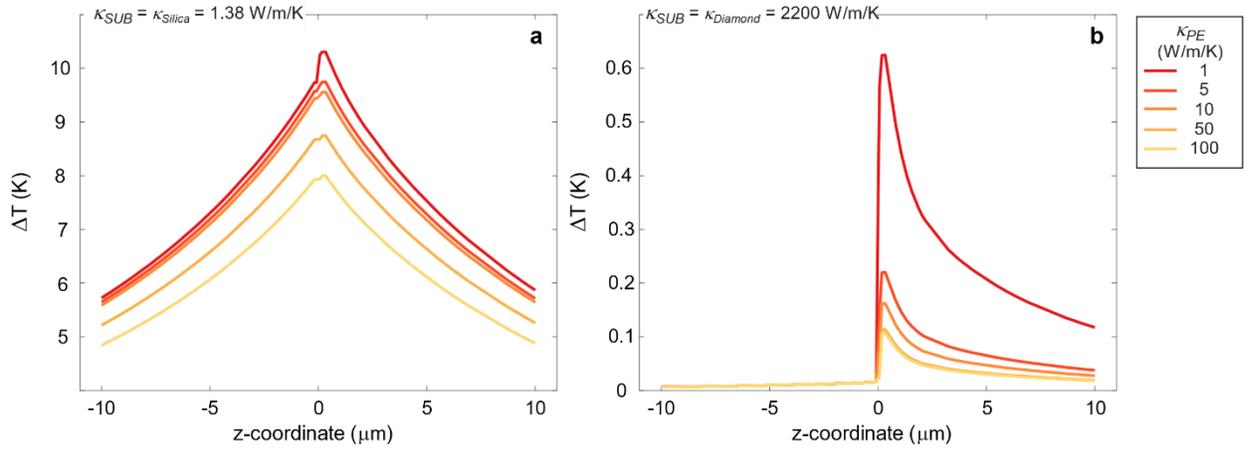

*Figure S2. Spatial profile of the simulated temperature increase $\Delta T$ evaluated along a vertical (z) line passing through the sample on a $\mu m$ range, i.e. from the substrate to the air above one of the central patches as in Figure 4b, for increasing value of the polymer sample conductivity, from 1 to 100 W/m/K. Substrate in (a) is silica ($\kappa$ = 1.38 W/m/K), in (b) it is diamond ($\kappa$ = 2200 W/m/K).*

Moreover, we complement the numerical results reported in Figure 4b (main text) with similar simulations performed by fictively varying the thermal conductivity, $\kappa_{PE}$, of the polymer layer within the cavity (the value used for the calculations shown in the main text for polyethylene was set to 0.38 W/m/K), in order to assess the impact of the polymer properties on the system thermal response. We thus computed the temperature increase along the sample vertical direction, $\Delta T(z)$, by considering $\kappa_{PE}$ spanning from 1 to 100 W/m/K (from dark to light color shades in Fig. S2), for both the same substrate as in the main simulations (silica, $\kappa_{SUB}$ =1.38 W/m/K, Fig. S2a) and a "heat-sink" substrate, featuring a much higher thermal conductivity (e.g. diamond, $\kappa_{SUB}$ = 2200 W/m/K, Fig. S2b).

By comparing the results of our calculations at different $\kappa_{PE}$, the role of the polymer conductivity is apparent. Higher values of $\kappa_{PE}$ entail a decrease of the thermal gradient across the polymer layer along the z-direction, as testified by the less pronounced and abrupt variation of the temperature around z=0 (i.e. at the PE center) when moving from dark- to light-shaded curves in Fig. S2. The same trend is predicted regardless of the structure substrate. However, when a "heat-sink" material is employed, the absolute temperature variation is lowered (compare vertical axes in Fig. S2a and S2b), with an almost negligible increase induced within such a highly conductive substrate.

### III. *Electromagnetic field patterns of the cavity modes*

A purely electromagnetic simulation is also carried out to map the electromagnetic fields buried within the optical nanocavity. The simulation in this case is an FDTD (finite difference in the time domain) calculation over an array with the same geometrical characteristics and the material properties as in the experiments. In particular, the geometry here simulates the optical response of a patch nanocavity with a side L=2⊠m, as in the T2NanoIR experiments shown in figure 5 (main text). A monitor recording the optical energy at different frequencies is placed parallel to the gold layers and at the centre of the polyethylene polymer, thus showing a plan-view of the fields inside the transducer material. Figure S3_leftPanel shows the electric fields (three upper color plots) for the x,y,z components and the magnetic fields (three lower color plots) for the same x,y,z components at the TM01 mode, which yields the peak experimentally found at 970cm$^{-1}$ in figure 2a (main text) and the intensity maps in figure 5a. Figure S3_rightPanel shows the same quantities but for the TM02 mode.

As expected, inside the optical nanocavity, i.e. in the concealed volume between the two upper and lower gold layers, the electric field is directed mainly along the z_component, while the x,y components are near to zero. Conversely, owing to the electric currents in the metals which build up the optical resonance, the magnetic field results to be mainly directed along the y-component. The resemblance of the $H_y$ components with the maps shown in figure 5 (main text), along with the considerations on the tip-sample interaction presented in the paper, let us speculate that the near-field coupling is enabled by the electric currents triggered by the AFM gold tip (see text for details).

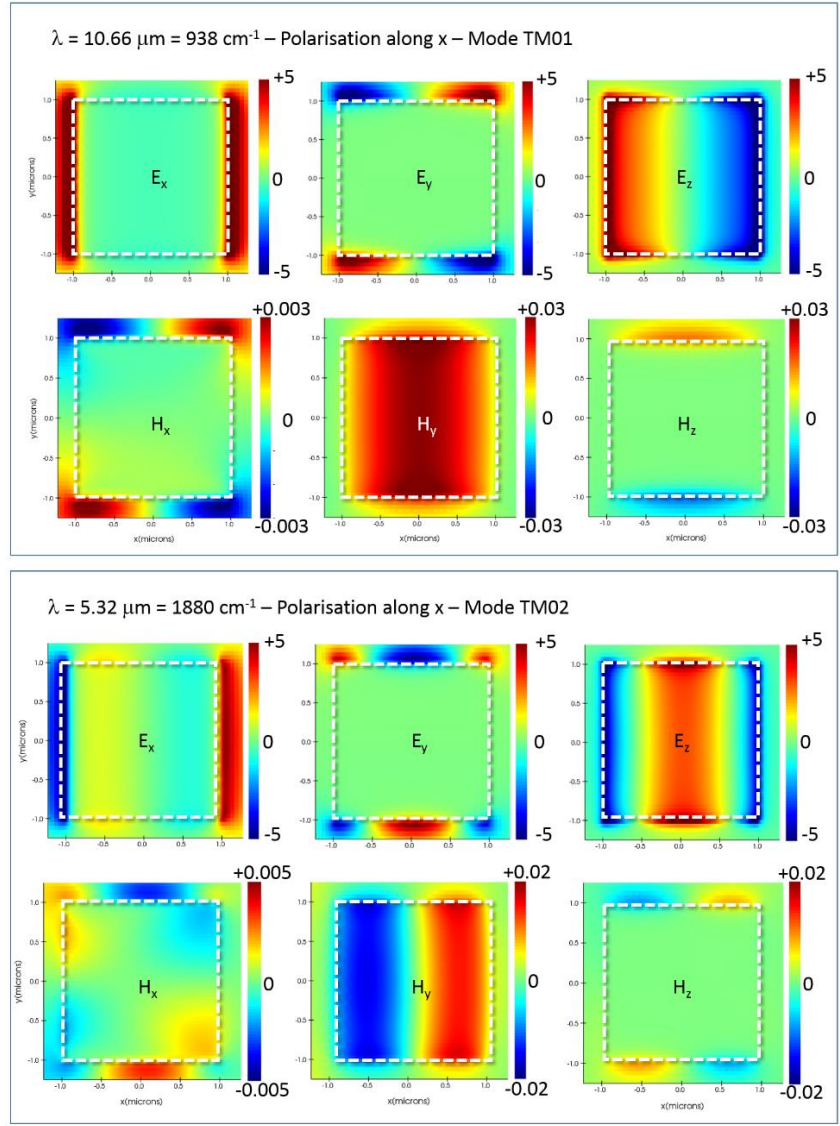

*Figure S3. FDTD numerical calculation of the x,y,z components of the electric field (upper color plots) and magnetic fields (lower plots) for the 1ˢᵗ resonance mode TM01 (left panels) and the 2ⁿᵈ resonance mode TM02 (right panels). The simulated geometry is the one that leads to the results of figure 5 in the main text: patch side L=2.0μm, non-doped sample.*

### IV.   Comparison of SNOM and T2NanoIR data on similar patches

We have collected on similar patch resonators SNOM and T2NanoIrR spectra on different positions on the top surface of the patches, marked with capital letters in the topographic AFM images in Figure S4 below. Both experiments have been performed with the light impinging from the right side of the images with the electrical field p-polarized (i.e. along the AFM tip shaft). One can notice at a glance that the SNOM spectra depend on the relative position of the tip on the resonator edge, due to a relative phase delay of the tip- and the patch- excitation. On the other hand, T2NanoIR spectra are insensitive to the position of the tip. In this sense, the presented approach yields unambiguous and easily interpretable results, albeit more sensitive to chemical contaminations and impurities.

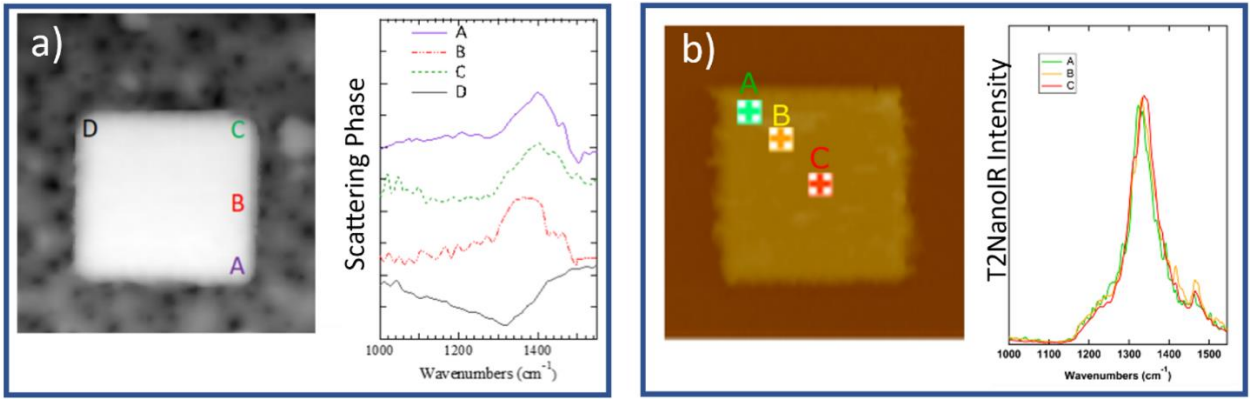

*Figure S4. a) Topographic AFM image of a patch antenna obtained in tapping mode with the NeaSNOM setup (NeaSpec) and relative Scattering Phase spectra collected at selected positions. b) Height image collected in contact mode in the NanoIR setup with spectra collected from one corner to the edge.*

### V.   Reproducibility of T2NanoIR spectra for different sample chips and within large antenna arrays

We have prepared with the aforementioned fabrication technique see Figure S1) other two samples with the same vertical architecture as those presented in the main text, although with a slightly different polymer thickness. All sample chips have the same doped QW structure as the one discussed in the main text. We have measured all both with FT-IR and T2NanoIR in order to detect and compare the anti-crossing behavior. On these samples we have performed measurements on both isolated antennas and in array and with both p- and s-polarized electrical field. The results are reported in the two panels of Figure S5. The small differences found in the three samples arise mostly from (i) different polymer thickness, (ii) small imperfections due to chemical impurities in the polymer layer and /or residuals from the fabrication process (iii) deviation from the perfectly squared shape (mainly for the smallest antennas, geometry around L=1μm). Overall, we find an excellent reproducibility within the obtained results.

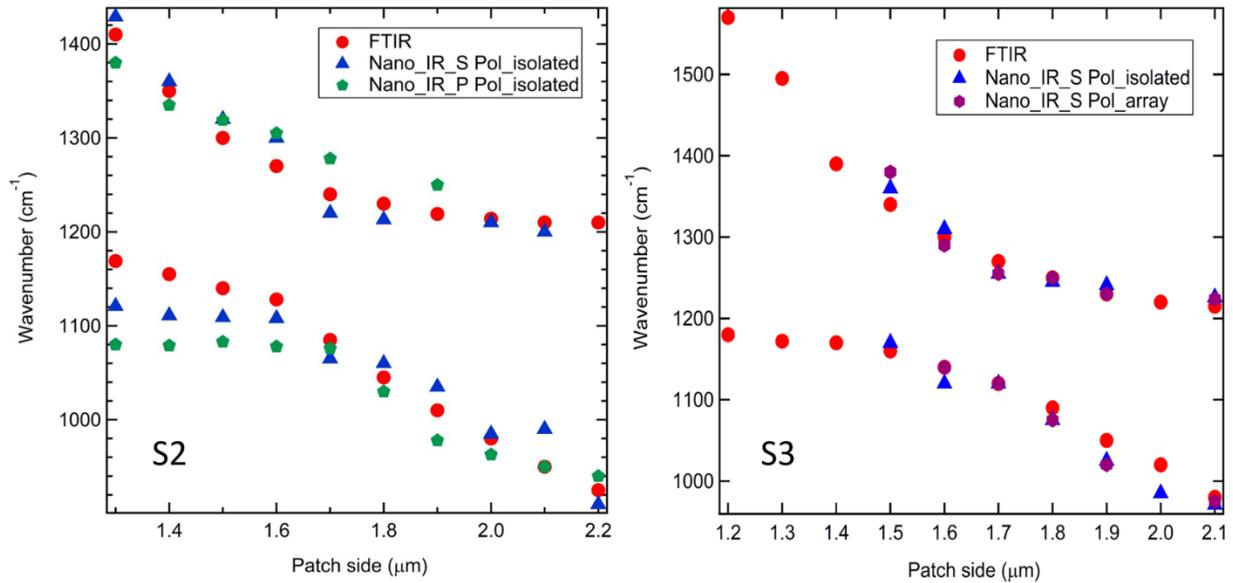

*Figure S5. Anti-crossing behavior as measured on sample S2 and S3 (sample S1 is the one described in the main text). The differences among the two samples resonant frequencies can be ascribed to different PE thickness in the sample, which produces a different effective index. For each sample we compare FTIR data with T2NanoIR spectra, in one case measured for both S and P polarization on isolated antennas (S2 sample) and in the other case for both isolated and array samples with S polarization.*

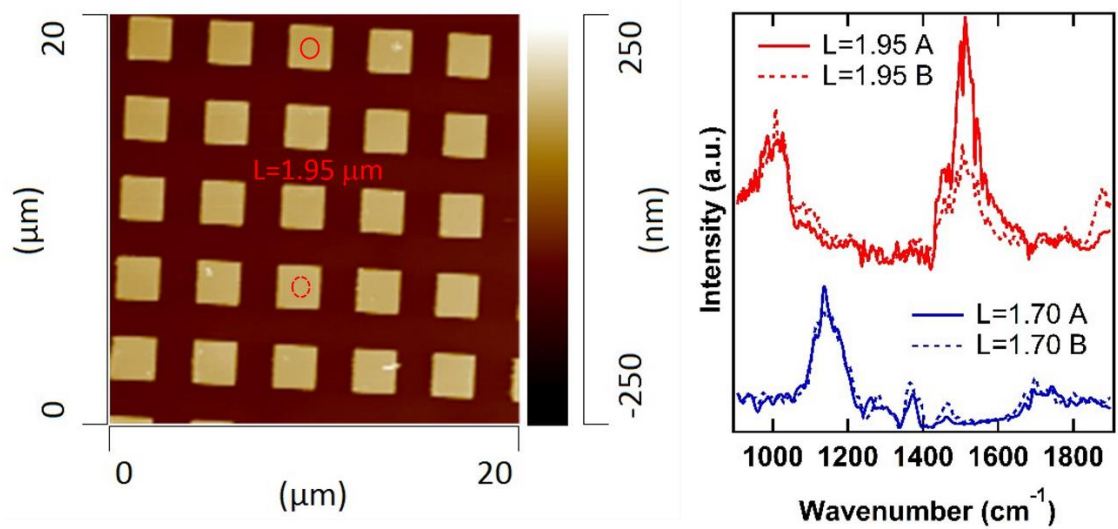

*Figure S6. Topographic height AFM image of different patch antennas in an array. In the plot on the right the spectra found on different antennas within the array (for two antenna side dimension, indicated in figure) are shown.*